\documentclass{ws-fixed}
\usepackage{subfigure}     
\usepackage{ws-rv-van}
\usepackage{graphicx}
\def\be{\begin{equation}}
\def\ee{\end{equation}}

\def\dd{\Delta^{(3)}}
\begin{document}

\chapter{Nuclear pairing:  basic phenomena revisited}

\author{G.F. Bertsch}
\address{Institute for Nuclear Theory and Dept. of Physics,
University of Washingtion, Seattle, Washington}

\begin{abstract}
I review the phenomena associated with pairing in 
nuclear physics, most prominently the ubiquitous presence of odd-even 
mass differences and the properties of the excitation spectra,
very different for even-even and odd-$A$ nuclei.  There are also
significant dynamical effects of pairing, visible in the inertias
associated with nuclear rotation and large-amplitude shape deformation.
\end{abstract}

\body

\section{Basic phenomena}

In this section I will present some of the basic manifestations of 
pairing in nuclei, using contemporary sources \cite{au03,bnl} for 
the experimental
data.  In later sections, I will describe in broad terms the 
present-day theoretical understanding of nuclear pairing, emphasizing
the many-body aspects rather than the aspects related to the 
underlying Hamiltonian.

\subsection{Pairing gaps: odd-even binding energy differences}
The basic hallmarks of pair condensates are the odd-even staggering in
binding energies, the gap in the excitation spectrum of even systems, and
the compressed quasiparticle spectrum in odd systems.  To examine odd-even
staggering, it is convenient to define the even and odd neutron pairing gaps
with the convention 
\be \dd_{o,Z}(N) = {1\over 2}(E_b(Z,N+1) - 2E_b(Z,N)+
E_b(Z,N-1)), \,\,\,{\rm for} \,\,N\,\, {\rm odd,} \ee \be \dd_{e,Z}(N) =
-{1\over 2}(E_b(Z,N+1) - 2E_b(Z,N)+ E_b(Z,N-1)), \,\,{\rm for} \,\,N\,\,\, {\rm
even.}
\ee 
where $N$ and $Z$ are the neutron and proton numbers and $E_b$ is the 
binding energy of the nucleus.  The proton
pairing gaps are defined in a similar way. With the above
definition, the gaps are positive for normal pairing.
The neutron pairing gaps are
shown as a function of neutron number in 
Fig. 1. The data for this plot was obtained from nuclear binding
energies given in the 2003 mass table \cite{au03}.
The upper panel shows the gaps centered on odd $N$.  Typically,
the odd-$N$ nuclei are
less bound than the average of their even-$N$ neighbors by about 1 MeV.
However, one sees that there can
\begin{figure}
\includegraphics[width = 11cm]{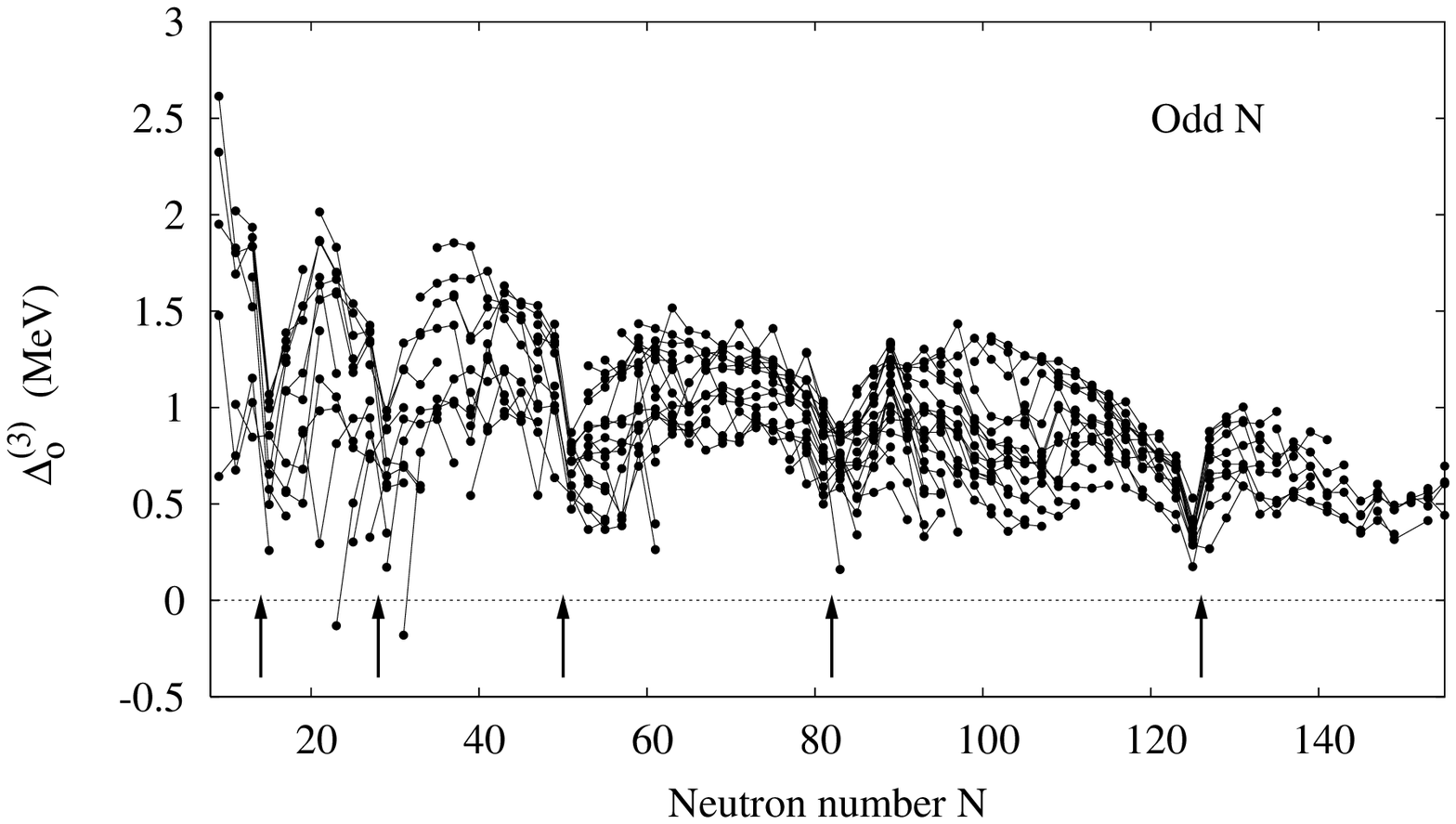}
\includegraphics[width = 11cm]{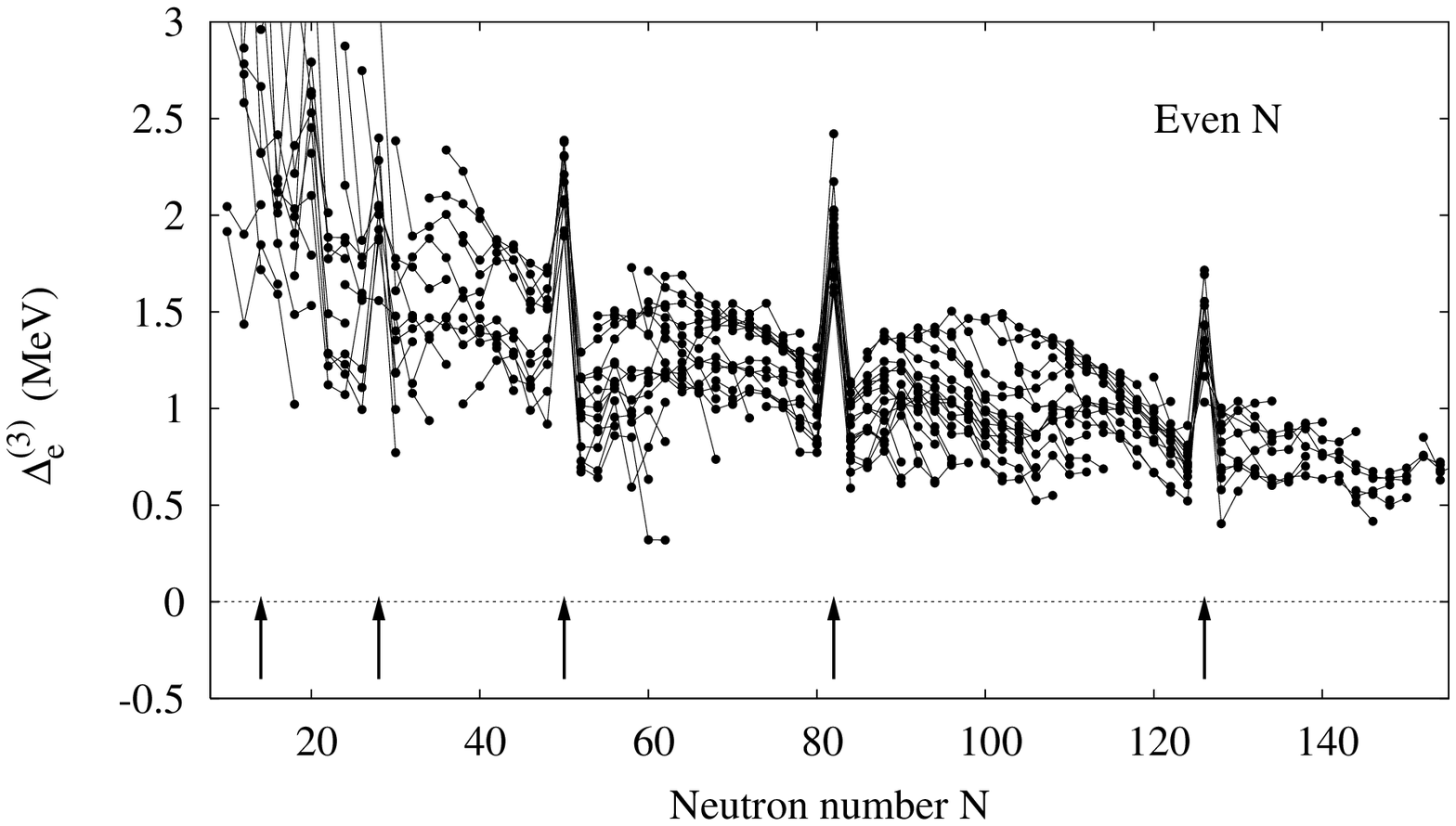}
\caption{Upper panels:  odd-N pairing gaps.  Lower panels: even-N
pairing gaps.}
\end{figure}
be about a factor of two scatter around the average value at a given
$N$.  
Note that there are two exceptional cases with negative $\dd$ for
odd neutrons, at $N=23$ and $N=31$.  I will come back them later.
One can also see a systematic trend in the gap values as a function 
of $N$, namely the gaps get smaller in heavier nuclei. I will
also come back to this behaviour in the theory discussion.
Another feature of the odd-$N$ gap systematics is the occurance of dips at
particular values of $N$.  In fact the dips occur adjacent to the 
well-known magic numbers $N=28,50,82$ and $126$. 
In addition there is a dip adjacent to $N=14$, which corresponds
to $n=2$ in the magic number sequence $\frac{1}{3}(n+1)(n^2+2 n+6)$.

The systematics of the even-$N$ gaps shown in the lower panel is 
similar with respect to the following: average values, the fluctuations at each
$N$, and the smooth trend downward with increasing $N$.  However,
the magic number anomolies are now very striking spikes that occur
exactly at the magic numbers.  Also, the average values in lighter
nuclei appear to be larger for the even-$N$ gaps than for the odd-$N$.
I will also come back
to this feature in the theory section.  

The corresponding systematics of proton gaps is shown in Fig. 2.  The
same qualitative features are present here as well, but the magic
number effects are less pronounced.  I do not know of any explanation
of this difference between neutron and proton pairing.    
\begin{figure}
\includegraphics[width = 11cm]{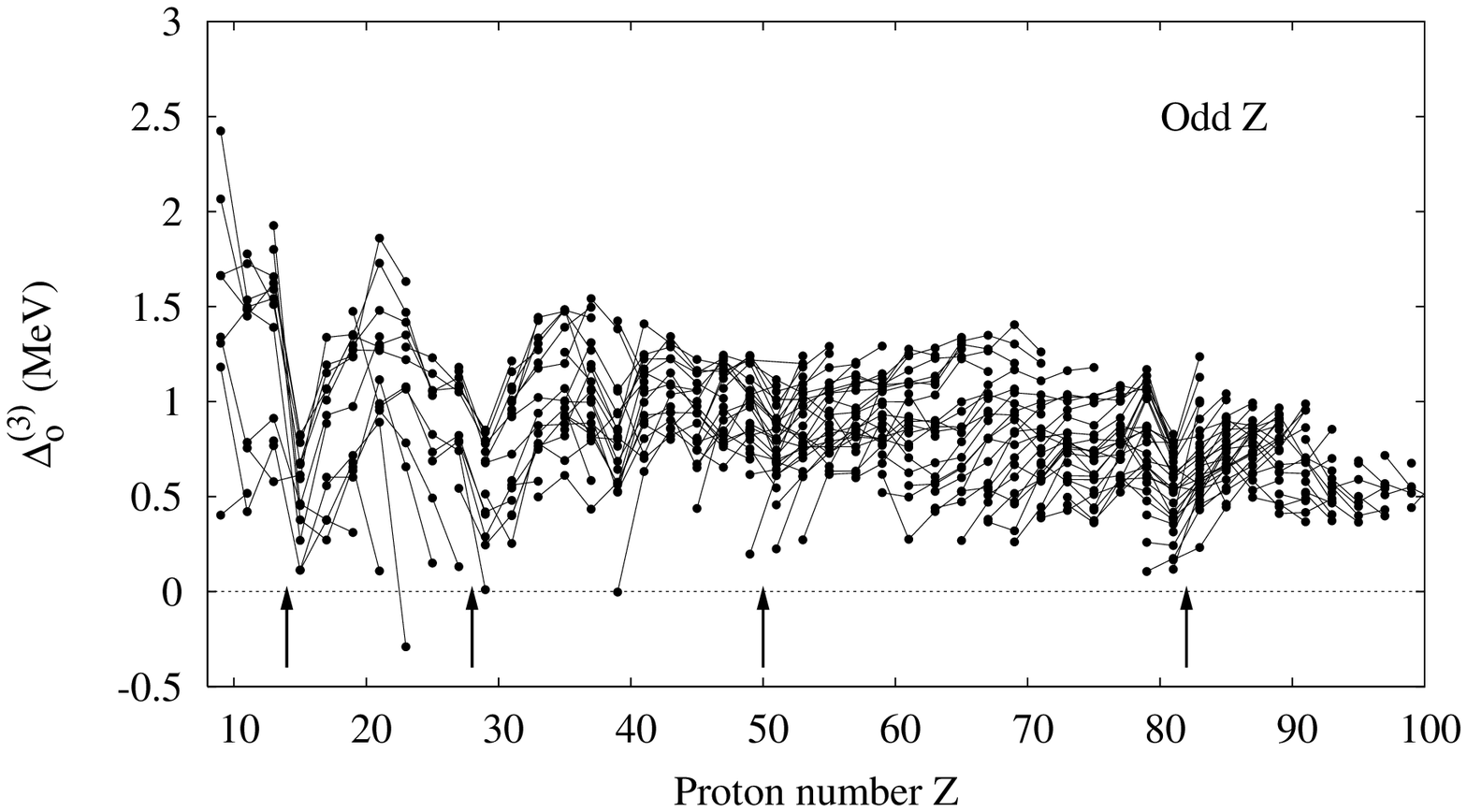}
\includegraphics[width = 11cm]{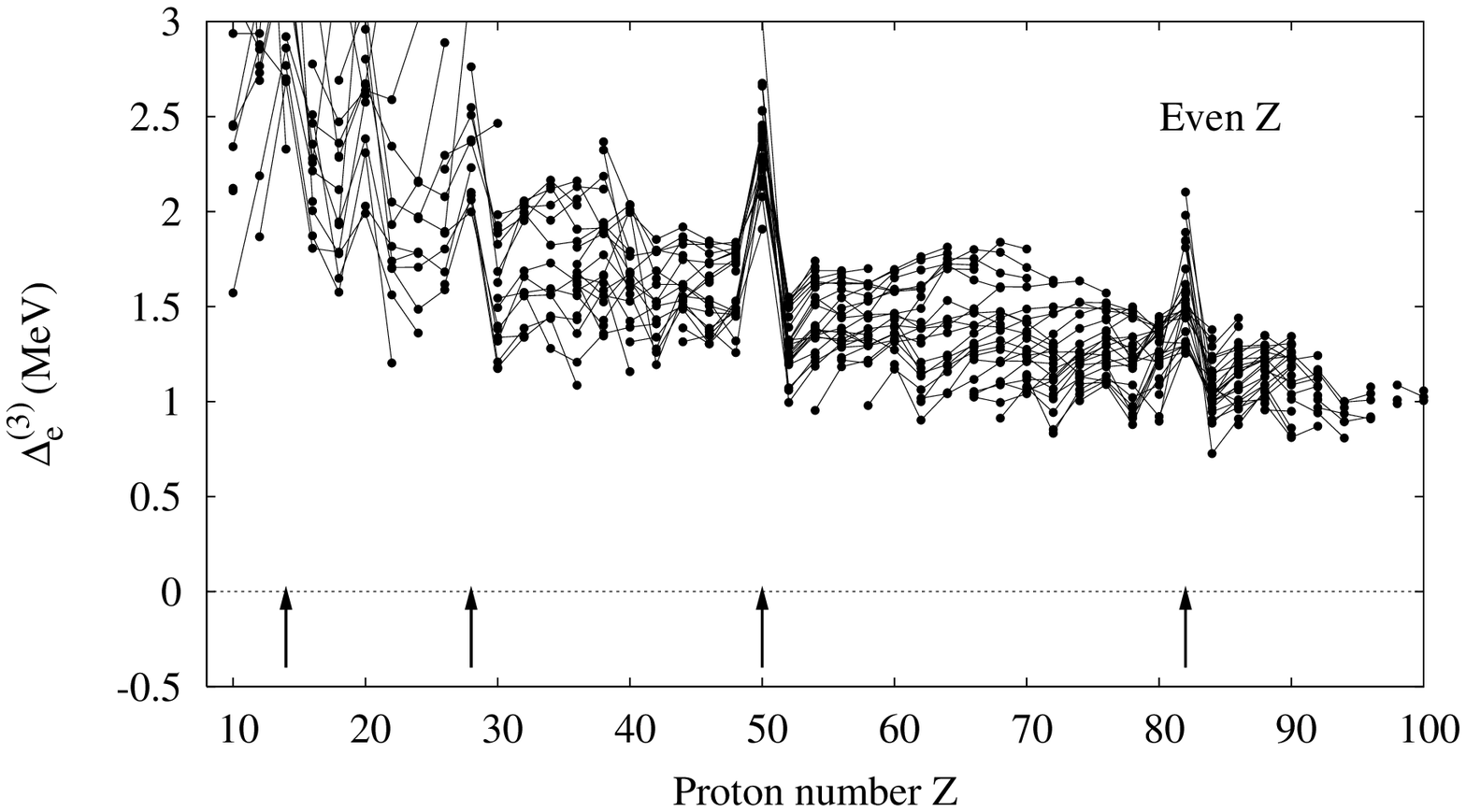}
\caption{Upper panels:  odd-$Z$ pairing gaps.  Lower panels: even-$Z$
pairing gaps.}
\end{figure}

The table below gives some fits to the pairing gap systematics.
Shown are the fitted values of the gap parameterizations and
the rms errors of the fits, in units of MeV.  
The simplest
model is a constant gap, $\Delta^{(3)} = C$, shown on the line
labeled $C$.  One sees that a typical gap size is 1 MeV, and typical
fluctuations about that are smaller by a factor of 3.  Beyond that,
there are differences between protons and neutrons and between the
odd and the even gaps.  The even gaps are somewhat larger and have
somewhat larger fluctuations, which is
to be expected in view of the shell effects exhibited in Fig. 1.  
The odd proton gap is smaller than the odd neutron gaps which might
be expected from the repulsive Coulomb contribution 
to the pairing interaction.  There is also a mean-field contribution 
of the Coulomb that has opposite signs for even and odd protons.  
Indeed the even proton gaps are actually larger
than their neutron counterparts.

For the next lines in the table, I come back to the broad trend
in Fig. 1, a systematic decrease in gaps with increasing mass number.
It is
conventional to describe this with a fractional power dependence,
$\dd = c/ A^{1/2}$.  This decreases the rms errors somewhat,
but there is no theoretical basis for
the fractional power of $A$.  In the last line I show the result
of a two-parameter fit to the functional form 
$\dd = c_1/A +c_2$.  This functional form is more justified by theoretical
considerations, as will be discussed in the theory section below.

\begin{table}
\begin{tabular}{|c|cccc|}
\hline
$\dd$ & protons & protons & neutrons & neutrons\\    
$o/e$   & odd & even & odd & even\\
\hline
data set  & 418 & 407 & 443 & 442\\
\hline
$C$ & $0.96\pm0.28$ & $1.64\pm0.46$ & $1.04\pm0.31$ & $1.32\pm0.42$\\ 
$c/A^{1/2}$ &  &  & $12/A^{1/2}\pm0.25$ & $12/A^{1/2}\pm0.28$\\ 
$c_1/A+c_2$ &  &  & $24/A+0.82 \pm 0.27$ & $41/A+0.94\pm0.31$\\ 
\hline
\end{tabular}
\end{table}

\subsection{Basic spectral properties} 

\begin{figure}
\includegraphics[width = 11cm]{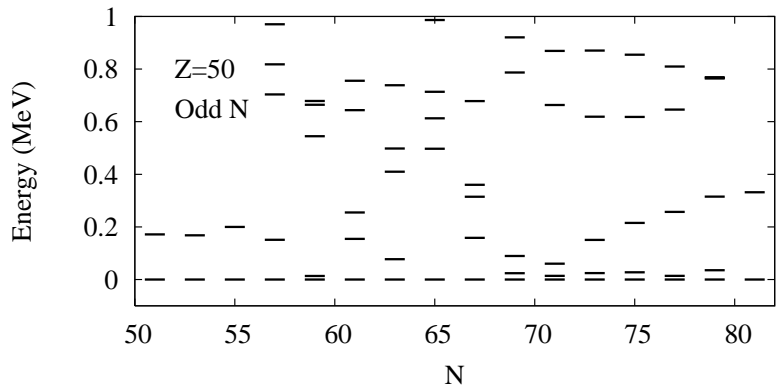}
\caption{\label{Sn} Energy levels of odd-$N$ Sn isotopes
}
\end{figure}

The other strong signatures of pairing are in excitation
spectra.  In the
simple BCS theory, the lowest excited states in an even system
requires breaking two pairs giving an excitation energy
\be
\label{E_ex}
E_{ex} \approx 2\Delta_{\rm BCS}.
\ee
On the other hand, in the odd particle number system, the quasiparticle
level density diverges at the Fermi energy.  This contrasting behavior
is very obvious in the nuclear spectrum.  As an example, the
isotope chain at proton magic
number $Z=50$ (the element Sn) has been a favorite for exhibiting and studying pairing
effects.  Figure \ref{Sn} shows the low-lying 
spectra of odd-$N$ members of the chain. 
One can
see that there are several levels within one MeV of the ground state.
The spins and parities of the levels (not shown) correspond very well 
with the single-particle orbitals near the neutron Fermi level.  However,
the spectrum is very compressed with respect to orbital energies calculated
with a shell model potential well. 
In contrast, the even members of the chain have no excited states at
all within the excitation energy range displayed in the Figure.

Let us look now look at the global systematics for excitations in
the even-even nuclei.  The estimate in Eq. (\ref{E_ex}) is too naive 
because there can be 
collective excitations within the gap, as is well-known from 
early days of BCS theory \cite{an58}.  For example, there
are longitudinal sound modes in an uncharged superfluid fermionic
liquid.  These have a phonon spectrum
allowing frequencies within the quasiparticle gap. 
One might expect that such modes would be absent in finite
systems when 
the size of the system is small compared to the coherence length
of the pairing field.  In fact  the situation for nuclear excitations
is much more complicated.  However, just for presenting the systematics,
we use the right-hand side of Eq. (\ref{E_ex}) to scale the 
excitation energies, taking the ratio
$E_2/2 \Delta$, where $E_2$ is the excitation energy and $\Delta$ is
the smaller of $\Delta_{eZ}^{(3)}$ and $\Delta_{eN}^{(3)}$.  

\begin{figure}
\includegraphics[width = 11cm]{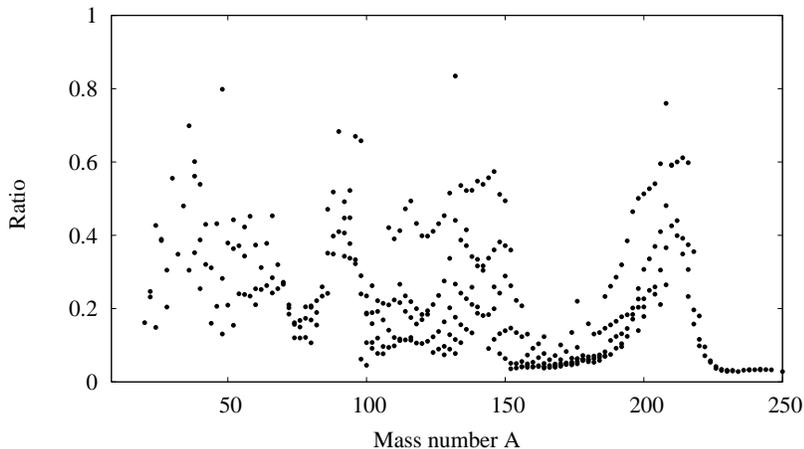}
\caption{\label{energy_gap} Energy gap in the excitation spectrum of even-even nuclei,
scaled to $2\Delta^{(3)}$.  See text for details.
}
\end{figure}

The scaled excitation energies of the first excited states in even-even 
are shown in Fig. \ref{energy_gap}.  With only a few exceptions these
states have
angular momentum and parity quantum numbers $J^\pi = 2^+$ and can
be considered to be collective quadrupolar excitations of the 
ground state. 
All of the ratios are smaller than one, with most in the range
0.1-0.5.  The very small excitations in the mass ranges 
$A=160-180$ and $220-250$ correpond to nuclei with static quadrupole
deformations.  

\begin{figure}
\includegraphics[width = 8cm]{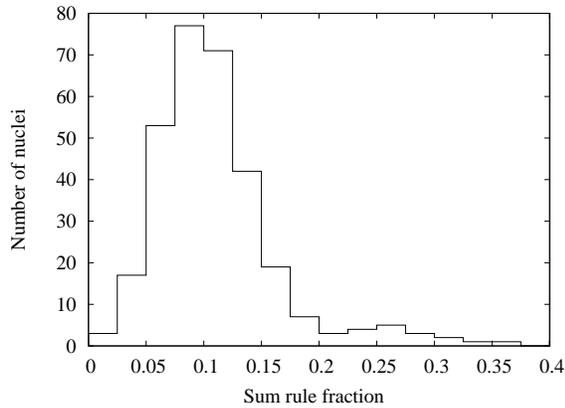}
\caption{\label{sr} Sum rule fraction for the first excited $2^+$ state in 
even-even nuclei.  See text for details.
}
\end{figure}

The physics underlying these excitations is the softness of a typical
nucleus with respect to quadrupolar
deformations.  On a qualitative level, the collectivity is similar 
to the phonon collectivity in the infinite Fermi gas.  A quantitative
measure of the collectivity is the sum-rule fraction contained in the
excitation, using the energy-weighted sum rule for some density operator.
For the phonon case, the
sum rule fraction approaches 100\% when the frequency of the collective
mode is small
compared to the gap \cite{an58}.  The collectivity in the
nuclear quadrupole excitations is quite different.  The sum
rule fraction carried by the lowest
$2^+$ excitation is more or less constant over the entire range
of nuclear masses, but it only about 10\% of the total (for
isoscalar quadrupole transitions.  This is known as the 
Grodzins systematics \cite{gr62,ra01}.  The observed distribution of
sum rule fractions is plotted as a histogram in Fig. \ref{sr}.

\begin{figure}
\includegraphics[width = 11cm]{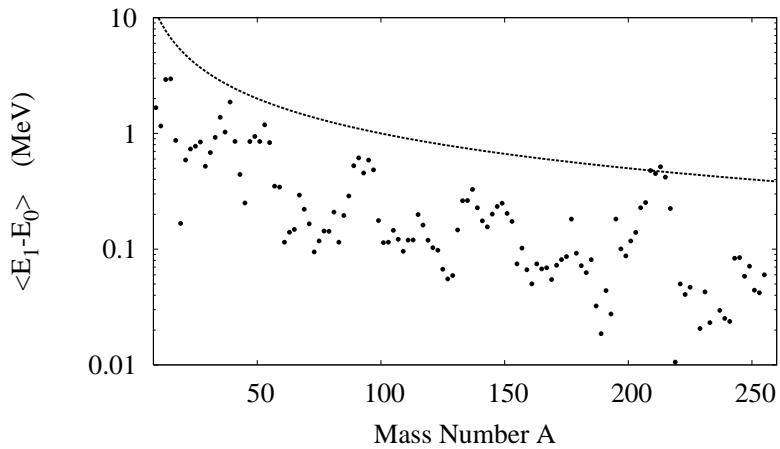}
\caption{\label{dnde}  Average energy of the first excited state
in odd-$A$ nuclei.  The dashed line is the Fermi gas estimate, Eq.   
\ref{eq:dnde}}
\end{figure}

Turning to odd-$A$ spectrao, some systematics related to the level
density are shown in Fig. \ref{dnde}.  The
average excitation energy of the first excited state is plotted
for each odd mass number $A$, averaging over even values of
$Z$.  For comparison, the solid line is the expected spacing
in the Fermi gas formula for the single-particle level density, 
\be
\label{eq:dnde}
{dn_s\over d E} = V{m k_F\over 2 \pi^2} \approx  {A\over 100} \,\,\,\,{\rm MeV}.
\ee  
The subscript $s$ on $N_s$ indicates that only one spin projection is counted,
and $k_F$ is the Fermi momentum.
$V$ is the volume of the nucleus, which is (roughly) proportional to the 
number of nucleons $A$.  One can see from the Figure that a typical spacing is a 
factor of 10 smaller than that given by the
Fermi gas formula.  Clearly interaction effects are at work
to increase
the level density near the ground state.

\section{Theory}

Mean-field theory has made enormous strides in nuclear physics; the
self-consistent mean field theory based on the Hartree-Fock-Bogoliubov
approximation and using semi-phenomenological energy functionals is
now the tool of choice for the global description of nuclear structure.
It is not my intention to review this subject since it is well covered
elsewhere in this volume.  

Neverless, there are number of aspects of nuclear pairing that can be
can rather easily understood using only the more qualitative aspects
of pairing theory.  Besides the pairing gaps and the effect on
level densities, there are important consequences for two-nucleon
transfer reactions and on dynamic properties
such as radioactive decay modes.  This section presents an overview
of some of these aspects.

\subsection{Mean-field considerations}

BCS pairing is not the only source of odd-even staggering in
binding energies.  As is well-known in the physics of finite
electronic systems, the Kramers degeneracy of single-particle
orbitals gives rise to an odd-even effect.  In a fixed potential
well, the pair-wise filling of the orbitals makes to a
contribution to $\dd_e$ that varies with system size as the
single particle level spacing, $\dd_e \sim A^{-1}$.  In
addition, the diagonal matrix elements of the two-body
interaction in the Hartree-Fock orbitals also contribute to the
odd-even staggering, both in $\dd_o$ and in $\dd_e$ \cite{du01}. 
In the nuclear context, the volume occupied by the orbitals is
(approximately) proportional to the mass number $A$, so this
interaction contribution also varies as $A^{-1}$.  

The last line of the Table shows a fit to the neutron pairing gaps
including an $A^{-1}$ term in the parameterization.  It does almost
as well as the phenomenological $A^{1/2}$ form.  Note also that
the coefficient of $A^{-1}$ is larger for the even gaps than
the odd ones.  This is just what is to be expected from the
contribution of the two-fold degenerate orbital energies. 

Another mean field effect can be interpreted by Eq. (5,6) below,
exhibiting the dependence of the pairing gap on the single-particle
level densities.  In general, level densities at the Fermi level
are higher in spherical nuclei than in deformed nuclei because of the
spherical shell degeneracy.  Thus, one expects larger pairing gaps
in spherical nuclei than in deformed.  Even more dramatic is the
shell quenching seen in the odd-$N$ gaps in Fig. 1.  The Fermi level
in the spherical nuclei showing quenched gaps turns out to be in the
$p_{1/2}$ or $s_{1/2}$ shell, which have low degeneracy.
Thus, the
occurance of the shell quenching is only partly due to the adjacent
magic number.

\subsection{Strength of the pairing interaction}
 
The BCS theory gives the following formula for the gap parameter \cite{sc64},
\be
\label{eq:BCS}
\Delta_{\rm BCS} = (E_{max}-E_{min}) \exp(-1/g)
\ee
where
\be
g= -G {d n_s\over d E}.
\ee
Here the prefactor of the exponential is the window of single-particle energies
for orbitals participating in the pairing and $G$ is the strength
of the pairing interaction.
Eq. (6) defines the dimensionless quantity $g$ that characterizes the
strength of the pairing condensate.

\begin{figure}
\includegraphics [width = 11cm]{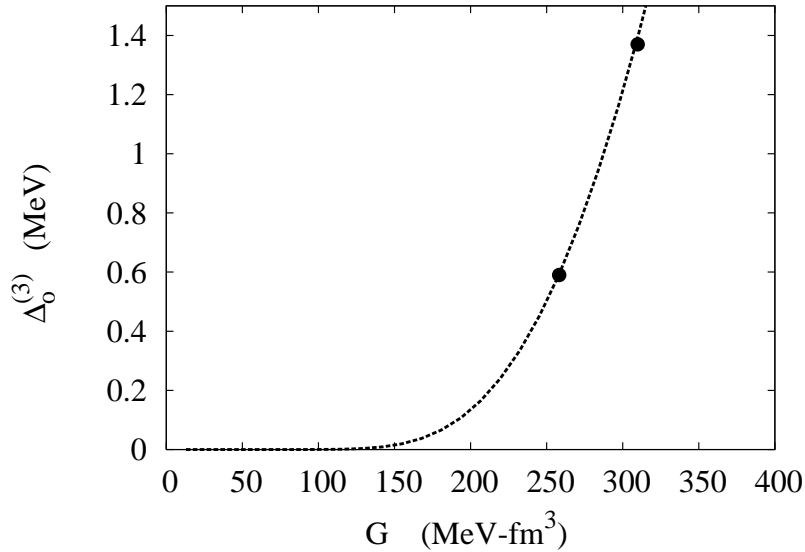}
\caption{\label{v0-dependence}
Filled circles: average pairing gaps for 443 odd-$N$ neutron gaps, 
calculated for two strengths of a contact pairing interaction
\cite{be09}.
Dot-dashed curve shows the dependence on
strength according to eq.  \ref{eq:BCS}.
}
\end{figure}

In present-day theory, the qualitative formula Eq. (\ref{eq:BCS}) is superceded by detailed
calculations of the orbitals and the pairing interaction, based on
Hartree-Fock (HF) mean-field theory or Hartree-Fock-Bogoliubov (HFB)  theory.
This permits the treatment of the interaction by
a two-nucleon potential and replacing of a generic level density by
computed single-particle level spectra.  However, there are significant
uncertainties about both these aspects, and the pairing interaction is
often parameterized in a simple way.  As an example, I show results
of a global 
study of pairing systematics that used the Skyrme energy functional for
the mean field and a contact interaction for the pairing \cite{be09}.  The 
odd-$N$ pairing gaps were calculated for two strengths of the
pairing interaction, giving average gaps shown as the filled circles in
Fig. \ref{v0-dependence}.
In the HFB calculations, the energy window was taken to be
$E_{max}-E_{min} = 100$ MeV.  Using this value in Eq. \ref{eq:BCS}, the
calculated average gap at $V_0/V_0^{sd} = 1$ is reproduced for
$g=0.20$.  Noting that $g$ depending linearly on the pairing strength,
Eq. \ref{eq:BCS} gives the dashed line as a function of $V_0$.  One
sees that there is a very strong dependence of the gap on the pairing
strength which is reproduced by the simple theory of Eq. \ref{eq:BCS}.
It is interesting to note also that Ref. [9]\cite{bo58} also 
estimated $g$ as $g \approx 0.2$ using the meager data available
at the time.

\subsubsection{Origin of the pairing interaction}

It is no surprise that conditions for pairing are satisfied in 
nuclei.
The nuclear interaction between identical nucleons 
is strongly attractive in the spin-zero channel, almost to the degree
to form a two-neutron bound state.  While this explains the origin
at a qualitative level, the many-body aspects of the nuclear interaction
make it difficult to derive a quantitative theory starting from 
basic interactions.  The progress one has made so far is reviewed in 
other chapters of this book, so I won't go into detail here.  But
just for perspective, I mention some of the major issues.  

I first recall problems with the mean-field interaction to use at
the Hartree-Fock level.
Most obviously, the effective interaction between nucleons in the
nuclear medium is strongly modified by the Pauli principle.  The
Pauli principle suppresses correlations between nucleons and that
in turn make the effective interaction less attractive.  Beyond that,
it seems unavoidable to introduce three-body interactions
in a self-consistent mean-field theory.  These interactions have two origins. 
The first
is the three-nucleon interaction arising from sub-nucleon degrees of 
freedom.  It has been convincingly demonstrated that such interactions
are needed to reproduce binding energies of light nuclei and to 
calculated the bulk properties of nuclear matter.  Besides this
more fundamental three-body interaction, there may an induced interaction
associated with the short-ranged correlations and their suppression in 
the many-body environment.  In the popular parameterization of the
effective interaction for use in mean-field theory, the three-body interaction energy 
has the same order of magitude as the two-body interaction energy.
It would thus seem to be a great oversimplication to ignore the 
three-body effects in the pairing interaction.

The last issue is the role of the induced interaction associated with
 low-frequency
excitations. We have seen
that the nucleus is rather soft to surface deformations.  the virtual 
excitation of these 
modes would contribute to the pairing in exactly the same way that
phonon provide an attractive pairing interaction for the electrons
in a superconductor.  The size of the induced interaction is estimated
in Ref. [10]\cite{br05}; it may well have the same importance as the
two-particle interaction.  Note that if low-frequency phonons were
dominant, the energy scale in Eq. (5) would be greatly reduced.  

\subsubsection{Spin-triplet pairing}

The strong attraction between identical nucleons was the starting
point for the discussion of the pairing interaction in the previous
section.  In fact, the attraction is even stronger between neutrons 
and protons in the spin $S=1$ channel. Here the interaction gives rise 
to the deuteron bound state.  Nevertheless, all the pairing phenomena 
seen above are a result of $S=0$ pairing between identical nucleons.  

This connundrum is resolved in two ways.  First of all, pairing
is only favored when all the particles can participate.  The
spin triplet interaction is only strong in neutron-proton pairs, so
it would be suppressed in nuclei with a large imbalance between
neutron and proton numbers.  The other factor working against
spin-triplet pairing is the spin-orbit field of the nucleus.  It
breaks the spin coupling of the pair wave function, but it is
more effective in the spin-triplet channel \cite{po98}. In any
case, an increase in nuclear binding energies is seen along the
$N=Z$ line, called the ``Wigner energy" \cite{lu03}.

\subsection{Dynamics}

The dynamic properties of an extended fermionic system depend
crucially on the presence of a pairing condensate, changing it
from a highly viscous fluid to a superfluid.  The effects in
nuclei are not quite as dramatic as in extended systems because
the pairing coherence length in nuclei exceeds the size of the
nucleus.  Nevertheless, the presence of a highly deformable
surface in nuclei requires that pairing be treated in a
dynamical way.

\subsubsection{Rotational inertia}

\begin{figure}
\includegraphics[width = 8cm]{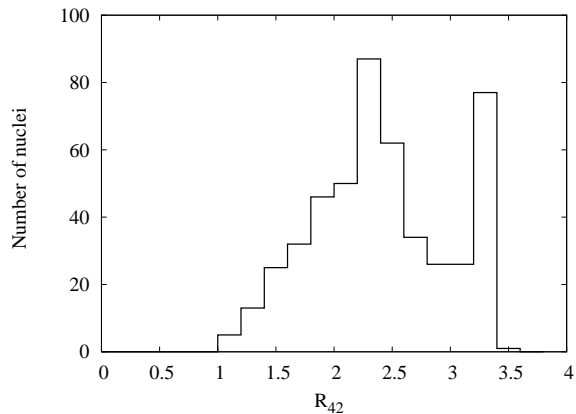}
\caption{\label{r42} Distribution of nuclei with respect to 
deformation indicator
$R_{42}$}
\end{figure}
The most clearly documented dynamic influence of pairing is its
effect on the moment
of inertia of deformed nuclei.  Without pairing, the rotational
spectrum of a deformed fermionic droplet is believed to follow 
the spectrum of a rigid rotor,
\be
E_J = {\hbar^2\over 2 {\cal I} } J (J+1).
\ee
Here $\hbar J$ is the angular momentum and the moment of inertia
${\cal I}$ would be close to the rigid value
\be
\label{eq:rigid}
{\cal I} \approx\frac{2}{3} A m \langle r^2\rangle\approx 
\frac{2}{5} A^{5/3} m r_0^2.
\ee
The author knows of no proof of this
assertion, but it can derived from the cranking approximation applied
to a many-particle wave function in a (self-consistent) deformed 
harmonic oscillator potential \cite[pp. 77-78]{BM2}. 
If the pairing were strong enough to make the
coherence length small compared to the size of the system, the
system would be a superfluid having irrotational flow and a
corresponding inertial dynamics.  What is somewhat surprising is 
that the weak pairing that is characteristic of nuclei still has
a strong effect on the inertia.

\begin{figure}
\includegraphics[width = 11cm]{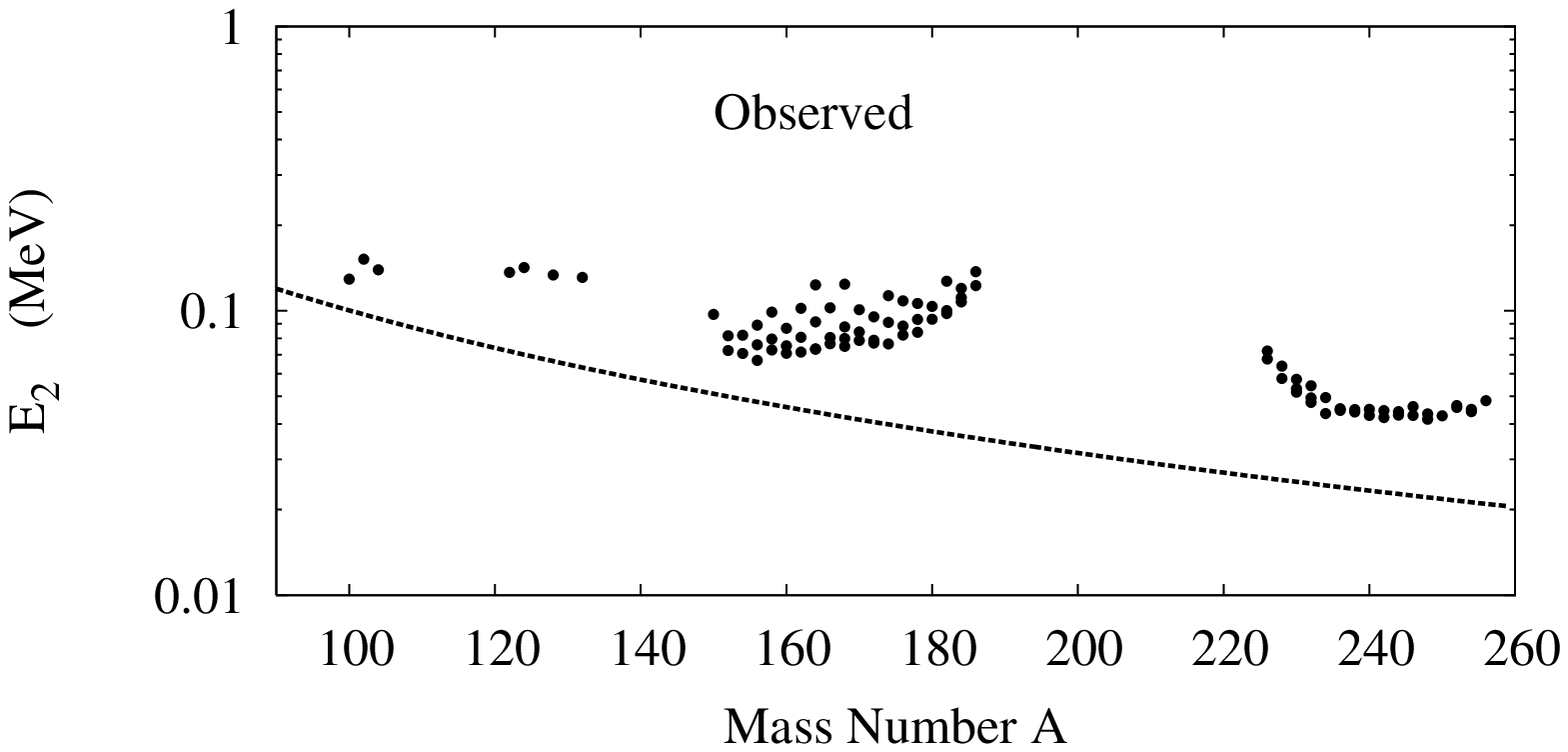}
\includegraphics[width = 11cm]{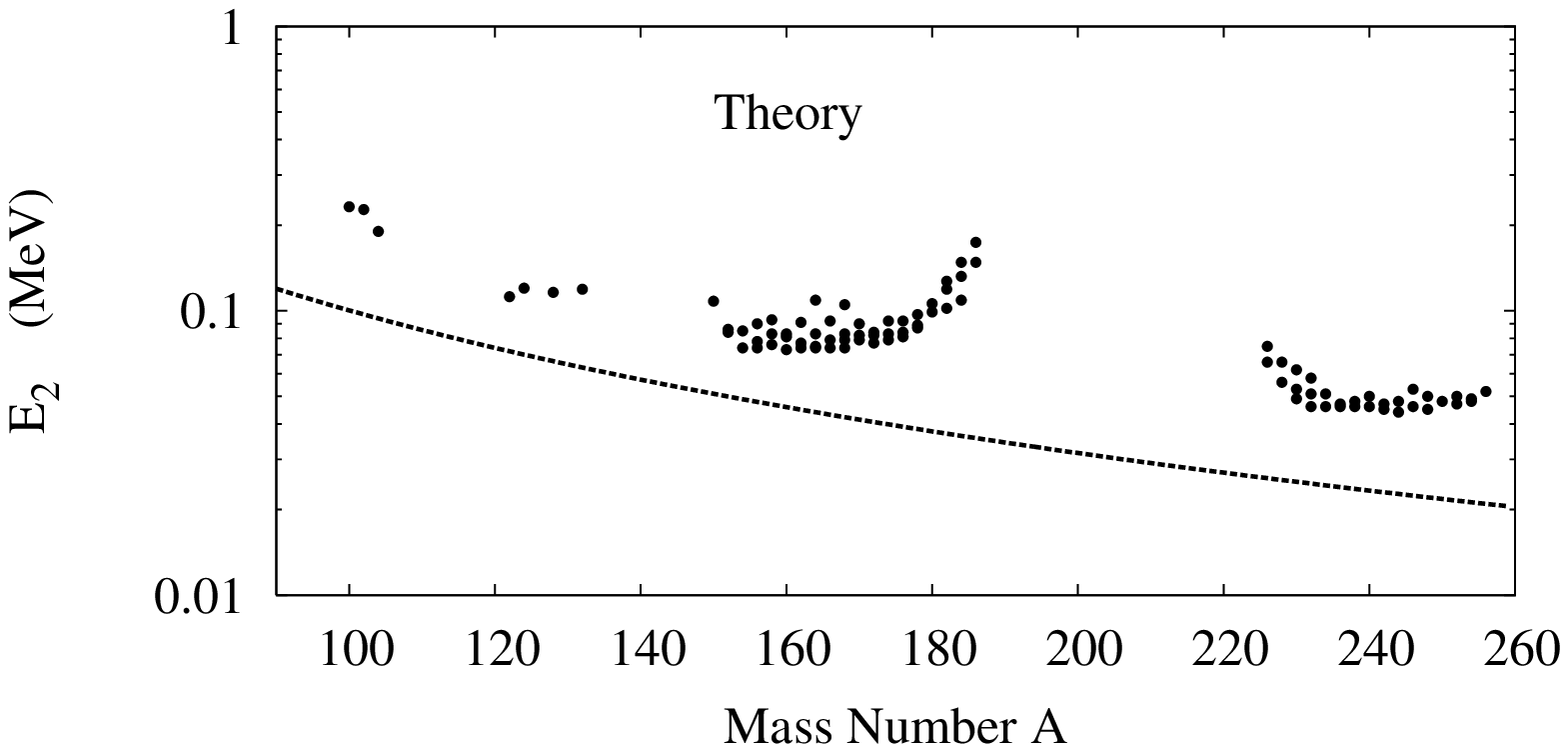}
\caption{\label{rigid}Excitation energy of the first $2^+$
state in deformed nuclei.  The line shows the prediction
assuming a rigid rotor. } 
\end{figure}

One can separate out the deformed nuclei from the others by making use
of the ratio excitation energies
\be
R_{42} = {E_4\over E_2}.
\ee
It is a good
indicator of the character of the nucleus and has the value 
$R_{42} = 10/3$ for an axial rotor.  A histogram of
$R_{42}$ for all the nuclei for which the energies are known is
shown in Fig. \ref{r42}.  There is a sharp peak around the
rotor value.  The $E_2$ excitation energies of the nuclei corresponding
to the peak are plotted in Fig. \ref{rigid} as a function of 
$A$. Also plotted (dashed line) is the predicted value assuming a 
rigid rotor, Eq. (\ref{eq:rigid}).
The experimental energies are systematically higher by a factor
$\sim 2 $, thus requiring inertias about half the rigid values.
Present-day self-consistent mean-field theory is very successful
in reproducing the experimental inertias, calculating them is
what is called the self-consistent cranking approximation.  As
an example, the lower panel of Fig. \ref{rigid} shows the calculated
$2^+$ energies using the HFB theory with an interaction that includes
pairing\cite{de10}.  The average energies
are very well reproduced, and the rms errors in the energies are only
$\pm10\%$.  While the theory works very well, it does not provide
a parametric understanding of the dependence of the inertia on
the pairing strength.  Naively one might have expected that
the effects would controlled by the ratio of the size of the
nuclei to the coherence length of the Cooper pairs, which is 
a small number.  We will also see in the next section another 
dynamic property showing a large influence of pairing.

\subsubsection{Large amplitude collective motion}

\begin{figure}
\includegraphics[width = 11cm]{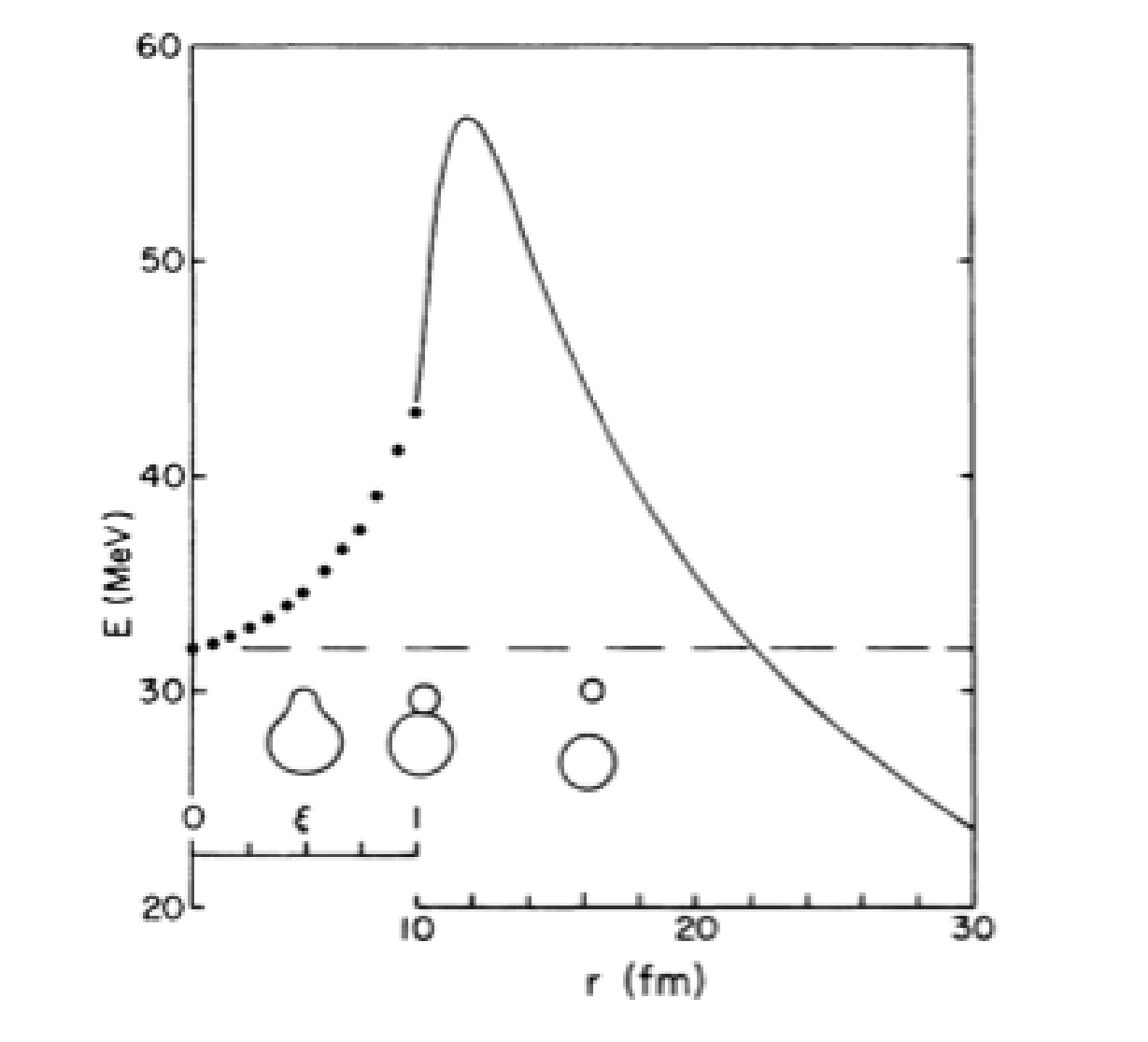}
\caption{\label{exotic} Potential energy curve for the decay
$^{223}$Rn $\rightarrow $ $^{209}$Pb + $^{14}$C. The outside
potential is a combination of Coulomb and nuclear heavy-ion
potentials. The dots show
the assumed Hartree-Fock states that connect the ground state
${223}$Rn configuration to the final-state cluster configuration.
}
\end{figure}
Also of interest, particularly in the theory of fission,
is the effect of pairing on large-amplitude shape
changes.  Qualitatively, it is clear that pairing promotes
fluidity.  The degree to which this happens can be examined in
one of the important observables of nuclear fission induced by
low-energy excitation, such as occurs in neutron capture.  The
observable is the internal energy of the fission fragments. 
With nonviscous fluid dynamics, the internal energy would be
largely deformation energy.  With more viscous dynamics, there
would be additional thermal energy.  So far, one has not been 
able to perform realistic enough calculations to compare theory
and experiment.  But the computational tools for the
time-dependent HFB theory are now reaching the point where such 
a test can be made.  (See Chapter X in this book).

Spontaneous fission is a decay mode that requires the nucleus to
tunnel under a barrier as it is changing shape.  This kind of
under-the-barrier dynamics is extremely sensitive to the
character of the system, whether it is normal or superfluid.  If
the system is normal, the relevant configurations under the
barrier are close to Hartree-Fock with relatively small
interaction matrix elements mixing different configurations.  On
the other hand, if there is pairing condensate, the interaction
between configurations can be enhanced by a factor
$2\Delta^2/G^2$ \cite[p. 159, Eq. (7.8)]{br05}, where here $G$ is
a typical interaction matrix element between neighboring mean-field
configurations.  Numerically, the pairing enhancement factor can
be an order of magnitude or more.
One should also keep in mind that in
tunneling, the lifetime depends exponentially on the inertial
parameters of the dynamics.  As an example, the nucleus 
$^{234}$U is observed to decay by many different channels,
ranging from alpha decay to spontaneous fission, and including
exotic modes such as emission of a Neon isotope.
The observed lifetimes of these decays range over 12
order of magnitude.  Theory including the enhancement factor is
able to reproduce the lifetimes to within one or two orders of
magnitude\cite[p. 163, Table 7.1]{br05}.  Without the
enhancement factor, there would be no possibility to explain
them.

\section*{Acknowledgment}

I wish to thank A. Sogzogni for access to the NNDA data resources.
I also thank A. Steiner and S. Reddy for helpful comments on the manuscript.
This work was supported by 
the US Department of Energy
under grant DE-FG02-00ER41132.

\bibliographystyle{ws-rv-van}
\bibliography{ws-rv-sample}

\end{document}